\documentclass[12pt]{iopart}
\usepackage{graphicx}
\usepackage{color}
\begin{document}

\title[Emergence of metallic surface states and negative differential conductance ...]{Emergence of metallic surface states and negative differential conductance in thin $\beta$-FeSi$_2$ films on Si(001)}

\author{Keisuke Sagisaka}
\address{Research Center for Advanced Measurement and Characterization, National Institute for Materials Science\\1-2-1 Sengen, Tsukuba, Ibaraki 305-0047, Japan}
\ead{SAGISAKA.Keisuke@nims.go.jp}
\author{Tomoko Kusawake}
\address{Research Center for Advanced Measurement and Characterization, National Institute for Materials Science\\1-2-1 Sengen, Tsukuba, Ibaraki 305-0047, Japan}
\author{David Bowler}
\address{London Centre for Nanotechnology, University College London, 17-19 Gordon St., London WC1H 0AH, United Kingdom}
\address{Department of Physics $\&$ Astronomy, University College London\\Gower St, London, WC1E 6BT, United Kingdom}
\address{International Centre for Materials Nanoarchitectonics (MANA), National Institute for Materials Science (NIMS), 1-1 Namiki, Tsukuba, Ibaraki 305-0044, Japan}
\author{Shinya Ohno}
\address{Yokohama National University\\79-5 Tokiwadai Hodogaya-ku, Yokohama, Kanagawa 240-8501, Japan}	
\ead{ohno-shinya-mv@ynu.ac.jp}

\vspace{10pt}
\begin{indented}
\item[]October 5, 2022
\end{indented}

\begin{abstract}
The electronic properties of the surface of $\beta$-FeSi$_2$ have been debated for a long while. We studied the surface states of $\beta$-FeSi$_2$ films grown on Si(001) substrates using scanning tunnelling microscopy (STM) and spectroscopy (STS), with the aid of density functional theory (DFT) calculations. STM simulations using the surface model proposed by Romanyuk \textit{et al.} [Phys. Rev. B 90, 155305 (2014)] reproduce the detailed features of experimental STM images. The result of STS showed metallic surface states in accordance with theoretical predictions. The Fermi level was pinned by a surface state that appeared in the bulk band gap of the $\beta$-FeSi$_2$ film, irrespective of the polarity of the substrate. We also observed negative differential conductance at $\sim$0.45 eV above the Fermi level in STS measurements performed at 4.5 K, reflecting the presence of an energy gap in the unoccupied surface states of $\beta$-FeSi$_2$.
\end{abstract}

%
%
%
%
%

\section{Introduction}
Thin films of $\beta$-FeSi$_2$ grown on a Si(001) substrate have generated considerable attention since the demonstration of intense infrared electroluminescence at $\sim$1.5 $\mu$m from a buried $\beta$-FeSi$_2$ epilayer on the Si(001) substrate was reported \cite{Leong1997}. In order to account for strong light emission from the epitaxial layer of $\beta$-FeSi$_2$, substantial efforts have been devoted to experimental and theoretical studies of the electronic structure of this semiconductor in both the form of a bulk single crystal and of a thin film, to address the issue of whether this semiconductor has a direct or indirect band gap \cite{Arushanov1995, Birdwell2002, Terai2011, Filonov1996, Clark1998, Moroni1999}.  Recently, it has been accepted that a single crystal of $\beta$-FeSi$_2$ has a direct band gap of 0.80-0.95 eV, and an indirect band gap of 0.7-0.78 eV \cite{Udono2004}, and that the epitaxial film remains an indirect gap semiconductor, even though it is strained by the lattice mismatch between $\beta$-FeSi$_2$ and silicon \cite{Birdwell2008}.

On the other hand, the electronic structure of the $\beta$-FeSi$_2$ surface is still debated: an early scanning tunnelling spectroscopy (STS) study showed a semiconducting surface state with a band gap of $\sim$0.9 eV \cite{Raunau1993}, whereas another STM study combined with X-ray photoelectron diffraction and ultraviolet photoelectron spectroscopy (UPS) showed a finite density of states (DOS) at the Fermi level, indicating a metallic surface state \cite{Hajjar2003}. This latter work suggested the formation of an $\alpha$-FeSi$_2$ film on the Si(001) substrate because of the observed metallic surface state, despite the fact that the surface structure and morphology of the film appeared very similar to those reported in Ref. 10. Moreover, Romanyuk \textit{et al}. recently experimentally confirmed the presence of a single crystalline thin film of $\beta$-FeSi$_2$ when they grew iron silicide on a Si(001) substrate \cite{Romanyuk2014}. They also proposed the presence of a metallic surface, as well as a bulk-like band gap ($\sim$0.85 eV) inside the film, based on the results of density functional theory (DFT) calculations, but without experimental data.  A general consensus on the electronic structure of this surface has not emerged despite two decades of study.

In this paper, we address the issue of the surface electronic structure of $\beta$-FeSi$_2$ ultrathin films grown on the Si(001) substrates from the results of precise STM and STS experiments, and DFT calculations. Our STS measurements detected a finite local density of states (LDOS) at the Fermi level, indicating that the reconstructed surface of $\beta$-FeSi$_2$ has a metallic surface state, which agrees well with our DFT calculations based on a $\beta$-FeSi$_2$/Si(001) slab model. Moreover, we observed negative differential conductance (NDC) at $\sim$0.45 eV above the Fermi level in the spectra. This phenomenon is reproduced well by a current-voltage (I-V) simulation using the surface electronic structure from our DFT calculations. The detection of NDC suggests the presence of a surface energy gap at the corresponding energy within the band gap of the $\beta$-FeSi$_2$ ultrathin film. Finally, we propose band diagrams for a $\beta$-FeSi$_2$ film grown on both n- and p-type Si(001) substrates estimated from the results of STS measurements. After discussing the methods used, we present STM and DFT results on the surface structure, before turning to the electronic structure.  We present differential conductance measurements and DFT simulations, and explore the band alignment and source of the negative differential conductance before concluding.

\section{Methods}
\subsection{Experimental method}
The experiments were carried out in ultrahigh vacuum (UHV) chambers with an STM apparatus (UNISOKU, USM-1400). The base pressure in the UHV chambers was kept below 4.0$\times$10$^{-9}$ Pa. The substrates used were pieces of n-type (P-doped, 0.01 $\Omega$cm) or p-type (B-doped, 0.013 $\Omega$cm) Si(001) wafers. After the substrate was well degassed for over 12 hours at 870 K in UHV, it was flashed to 1420 -- 1470 K for 10 -- 20 s several times to obtain a clean surface. Subsequently, iron was deposited on the clean surface at room temperature (RT) with an e-beam evaporator (AVC, AEV-3) and the sample was annealed to the temperature of 730 K for 10 min. According to a cross-sectional analysis by transmission electron microscopy (TEM), this operation produced a $\beta$-FeSi$_2$ film with a thickness of 1.0 -- 2.0 nm  \cite{Sagisaka20}. STM/S measurements were performed at the sample temperature of 78 K or 4.5 K. For the probe, chemically etched tungsten tips were used. All STM images were generated by the WSxM software \cite{Horcas07}.

\subsection{Computational method}
DFT calculations were performed with the VASP code \cite{kresse93, kresse96}. We employed the projector augmented wave (PAW) method with a plane wave cutoff of 500 eV, and used the generalized gradient approximation with the Perdew-Burke-Ernzerhof (PBE) exchange correlation functional\cite{Perdew96}. The $\beta$-FeSi$_2$/Si(001) surface was modeled using a periodic slab consisting of a 2$\times$2 surface with 16 layers of $\beta$-FeSi$_2$ stacked on 21 layers of silicon and a vacuum thickness of 19 \AA\ (see figure \ref{figlabel_A1}). The crystal structure of bulk $\beta$-FeSi$_2$ is shown in Fig.~\ref{figlabel_A1}(a)\cite{Dusausoy71}.  The top layer of the slab was terminated with silicon, according to a recent report on the structural determination of the $\beta$-FeSi$_2$ film by a low energy electron diffraction (LEED)-IV measurement \cite{Romanyuk2014}, and the bottom silicon layer was terminated by hydrogen atoms. The optimized PBE lattice constant for bulk Si in our calculation was 5.47 \AA\, leading to a surface unit cell length of 7.73 \AA. Experimentally, the $\beta$-FeSi$_2$ layer is observed to be pseudomorphically strained to the substrate; the interface structure between $\beta$-FeSi$_2$ and Si(001) was precisely determined by an analysis of the cross sectional TEM images \cite{Sagisaka20}.  We also checked that the impact of the interface structure on the surface electronic structure is negligibly small for the thickness of the iron silicide film used in the present study.
To match the substrate, the lattice constant of $\beta$-FeSi$_2$ (a = 9.86 \AA, b = 7.79 \AA, and c = 7.83 \AA) [figure \ref{figlabel_A1}(a)]\cite{Dusausoy71} needed to be adjusted in the plane parallel to the surface (b and c axes). As a result, the lattice of $\beta$-FeSi$_2$ was compressed by 0.8 \% and 1.3 \% in the b and c axes, respectively.
Structural relaxations were performed with 8$\times$8$\times$1 Monkhorst-Pack mesh Brillouin zone (BZ) sampling centered at the $\Gamma$ point until the force on each atom was reduced less than 0.02 eV/\AA. Calculations of LDOS and charge density were carried out with 22$\times$22$\times$1 BZ sampling. After the electronic ground state was calculated by the VASP code, STM images and tunnelling spectra were generated with the Tersoff-Hamann method \cite{Tersoff83, Tersoff85} implemented in the bSKAN code \cite{Hofer03}.

\section{Results and discussion}

\begin{figure}
	\begin{center}
		\includegraphics[width=8.0cm]{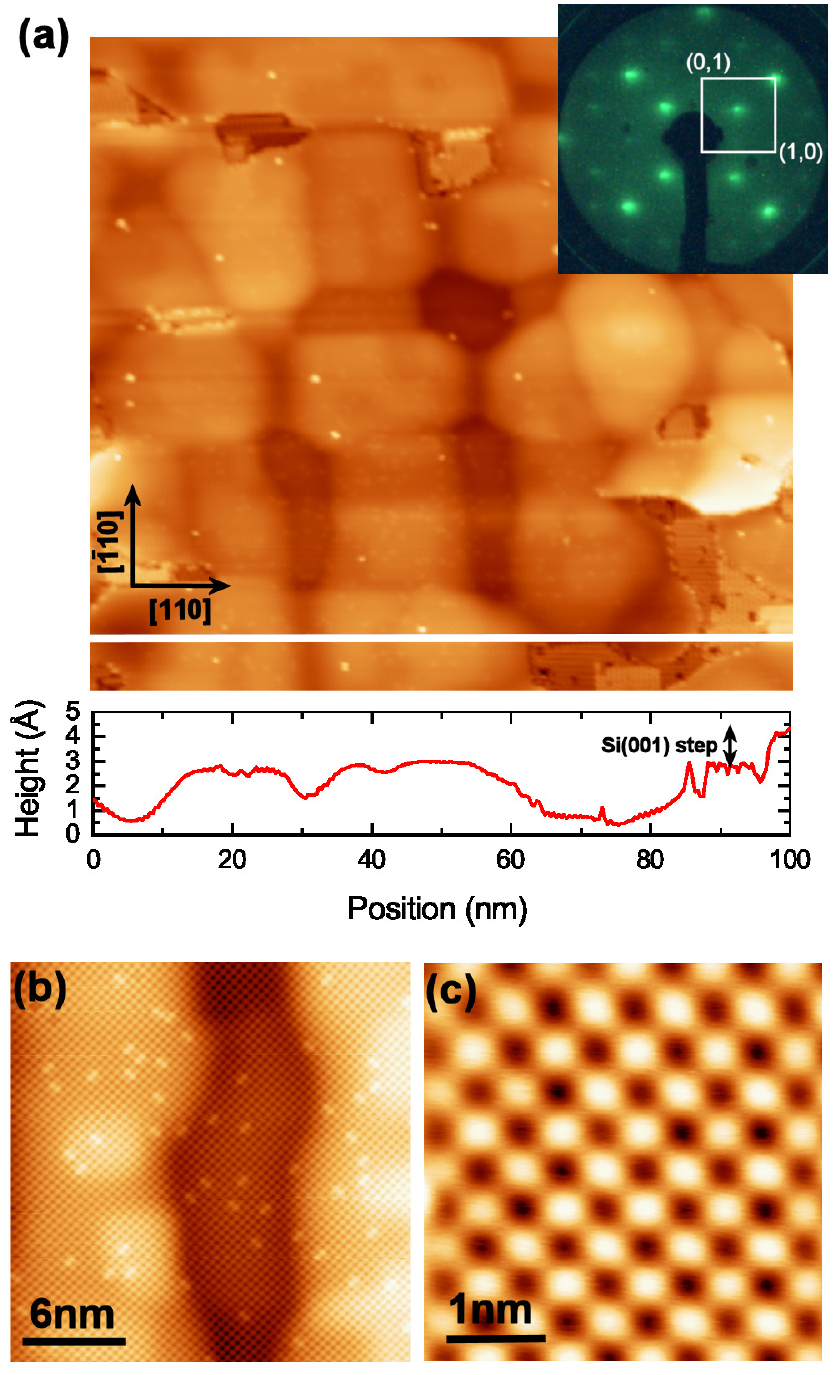} 
	\end{center}
	\caption{STM images of a $\beta$-FeSi$_2$ film grown on the Si(001) substrate. (a) Image size = 100 $\times$ 100 nm, Sample bias ($V_s$) = -1.5 V, tunnelling current ($I$) = 50 pA. Crystal orientations are denoted with respect to those of silicon substrate. The topographic profile was obtained from the white line crossing the $\beta$-FeSi$_2$ region and a small residual fraction of the Si(100) surface in the STM image. Inset: LEED pattern from the sample; incident electron energy = 67.5 eV, (b) Image size = 30 nm $\times$ 30 nm, $V_s$ = -1.5 V, $I$ = 10 pA, and (c) Image size = 5 $\times$ 5 nm, $V_s$ = +1.0 V, $I$ = 50 pA. Sample temperature ($T_s$) = 78 K.
	}
	\label{figlabel_1}
\end{figure}

Figure \ref{figlabel_1}(a) shows a typical STM image of the ultrathin iron silicide film covering almost the entire surface of the substrate. The surface appears to be rumpled, with a pattern of patches each several nanometres wide, with different image contrast: the root mean square surface roughness in the image was estimated to be 0.129 nm [see the cross sectional profile in figure \ref{figlabel_1}(a)]. However, the surface lattice is continuous and the surface profile smoothly varies without steps across the boundaries of the patched areas [figure \ref{figlabel_1}(b)]. According to the result of our TEM observations \cite{Sagisaka20}, this film consisted of a single crystalline $\beta$-FeSi$_2$ whose surface was smooth, but rumpled; the underlying stepped structure of the Si (001) substrate was compensated by steps in the $\beta$-FeSi$_2$ film at the interface, leading to a variation of the thickness between three values: 1.0, 1.5 and 2.0 nm. These values are close to the unit cell size of $\beta$-FeSi$_2$ normal to the (100) plane ($a$ = 9.86 \AA), 3/2$a$ and 2$a$, respectively. The crystallinity of the surface over a wide area was also confirmed by LEED measurements, which show slightly diffuse (1/2, 1/2) and (1, 1) spots [inset of figure \ref{figlabel_1}(a)]. We note that this spot spread is derived from the patches on the surface observed by STM, and it suggests that the lattice constant along the surface fluctuates slightly from patch to patch, or along the boundaries between adjacent patches. Zoomed STM images clearly show protrusions with c(2 $\times$ 2) periodicity as shown in figures \ref{figlabel_1}(b) and \ref{figlabel_1}(c). These characteristics are consistent with the results of a recent study on the $\beta$-FeSi$_2$(001) film using LEED $I-V$ analysis and DFT calculations \cite{Romanyuk2014}.  

\begin{figure}
	\begin{center}
	\includegraphics{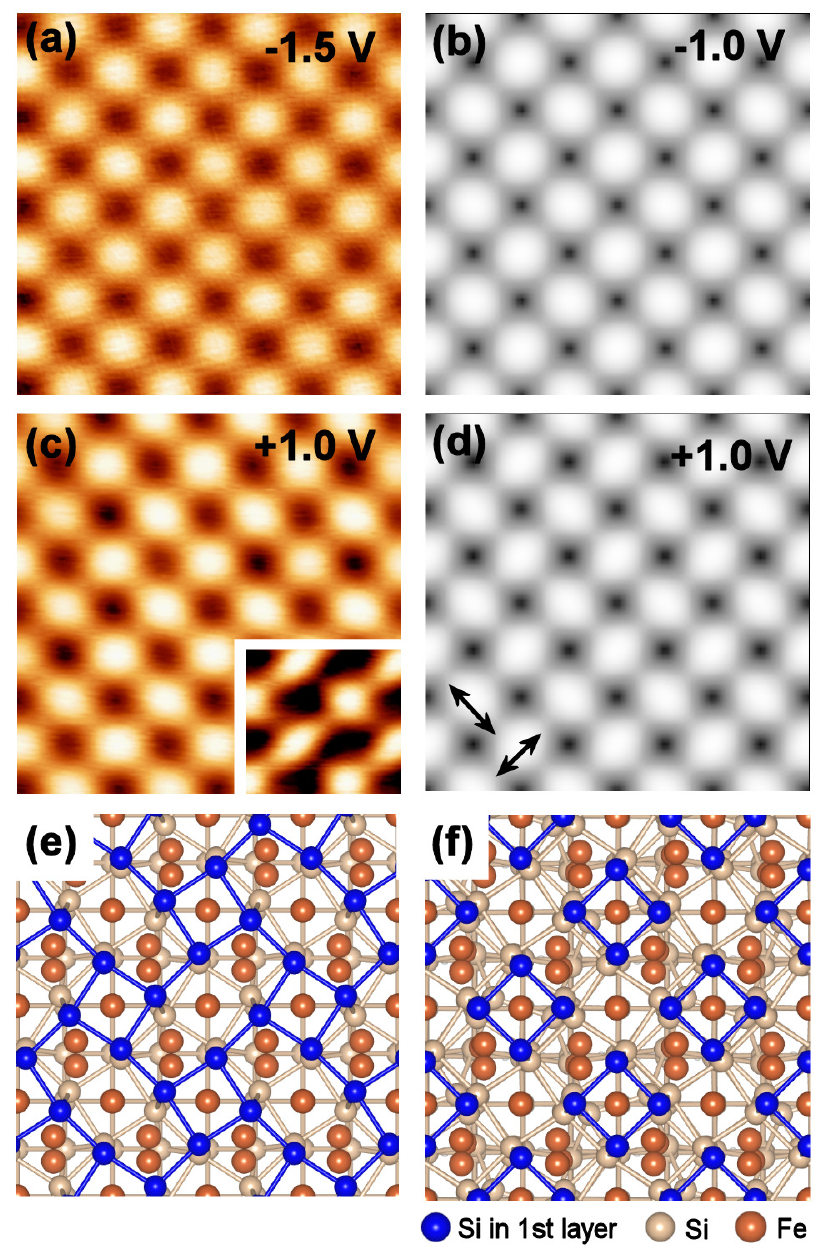}
	\end{center}
	\caption{Experimental [(a) and (c)] and simulated [(b) and (d)] STM images of the $\beta$-FeSi$_2$ surface. $V_s$ = (a) -1.5 V, (b) -1.0 V, (c) 1.0 V, and (d) 1.0 V. The image contrast is enhanced in the inset in (c) to emphasise the diagonal feature. (e) Top view of the bulk truncated $\beta$-FeSi$_2$(100) surface terminated with silicon atoms and (f) reconstructed surface after structural optimization, with which STM simulation was conducted.}
	\label{figlabel_2}
\end{figure}

The model proposed to explain the emergence of c(2 $\times$ 2) periodicity in STM images \cite{Romanyuk2014} proposed that four silicon atoms on the bulk-truncated surface [figure \ref{figlabel_2}(e)] cluster to form a tetramer [figure \ref{figlabel_2}(f)]. The Si tetramers are arranged with c(2 $\times$ 2) periodicity and are imaged as single protrusions by STM. Using this Si-tetramer model, our DFT-based simulation reproduced the experimental STM images in detail, at both positive and negative bias. Figure \ref{figlabel_2} displays a comparison of the experimental [(a) and (c)] and simulated [(b) and (d)] STM images of the $\beta$-FeSi$_2$ surface. Our DFT calculations found that the reconstruction of the surface reduces the energy by 0.275 eV per Si atom. 
A careful inspection of both experimental and simulated images reveals that each Si cluster is actually elongated alternately along (100) and (010) crystal orientations at a positive sample bias (unoccupied state), while appearing almost circular at a negative sample bias (occupied state).  This alternating elongation is probably derived from the atomic arrangement of Si atoms in the fourth layer [Si(2) in figure \ref{figlabel_A2}] and Fe atoms in the fifth layer [Fe(2) in figure \ref{figlabel_A2}] from the surface.

\begin{figure*}
	\begin{center}
	\includegraphics[width=16cm]{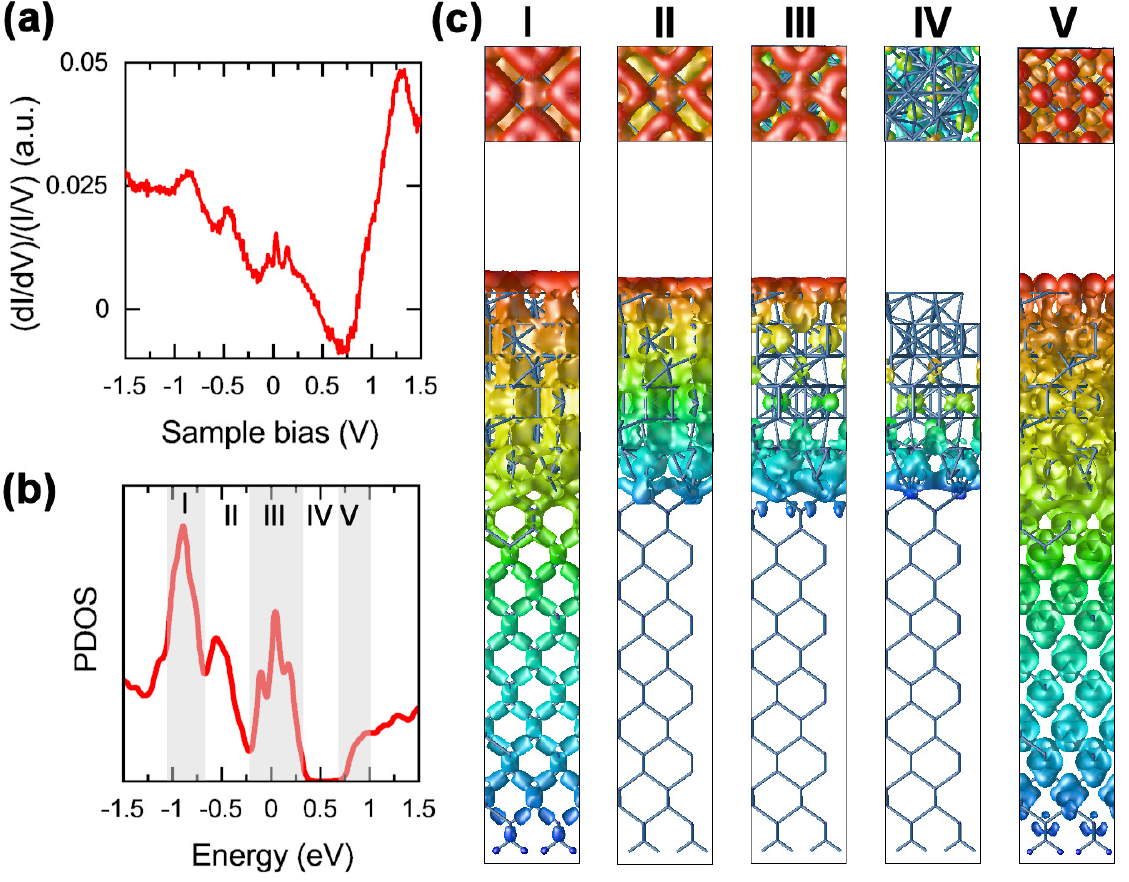}
	\end{center}
	\caption{(a) Normalized $dI/dV$ spectrum recorded on the $\beta$-FeSi$_2$ surface at 4.5 K. $dI/dV$ measurement was performed by placing the STM tip over a Si tetramer under open-feedback through a lock-in technique with bias modulation of 10 mV at a frequency of 830 Hz. Set point : V$_s$ = 1.5 V and I = 100 pA. Ten spectra were averaged. (b) Projected density of states (PDOS) for the top Si atom of the $\beta$-FeSi$_2$ slab model, (c) Isosurface map of charge density across the entire slab integrated in energy windows of (I) -1.1 $\sim$ -0.70 eV, (II) -0.70 $\sim$ -0.23 eV, (III) -0.23 $\sim$ +0.30 eV, (IV) +0.30 $\sim$ +0.7 eV, and (V) +0.70 $\sim$ +1.0 eV. Isosurface value: 10$^{-2}$ e/\AA$^3$. Color scale represents the $z$ position in the slab.}
	\label{figlabel_3}
\end{figure*}

Figure \ref{figlabel_3}(a) shows a typical normalized $dI/dV$ spectrum acquired from the $\beta$-FeSi$_2$ surface at 4.5 K. The flat and clean region of the surface did not have any distinct site dependence. A significant feature in the STS is the presence of finite conductance around the Fermi level, with three distinct peaks. Therefore, we conclude that the clean surface of the $\beta$-FeSi$_2$ has no energy gap, and so is metallic. This result is consistent with a previous report of a finite DOS of states at the Fermi level observed by UPS measurement \cite{Hajjar2003}, where the surface was assigned to be $\alpha$-FeSi$_2$, but actually it was $\beta$-FeSi$_2$. We also note that a semiconducting feature observed in the I-V curve reported by Raunau \textit{et al}. \cite{Raunau1993} was probably due to an experimental malfunction such as contamination of the STM tip. Another intriguing feature in our spectrum is a conductance drop to negative values above the sample bias of +0.45 V (NDC) which is discussed fully below.

To find the electronic structure responsible for the peaks detected in STS, we examined both the projected density of states (PDOS) for the top surface Si atoms, and the spatial distribution of electronic states in specific energy windows, using the slab model shown in figure \ref{figlabel_A1} (we note that the PDOS does not correspond directly to the normalised differential conductance, but gives more insight into the origin of different features; we compare experimental and theoretical differential conductance in figure~\ref{figlabel_5} below).
The peaks in the calculated PDOS plotted in figure  \ref{figlabel_3}(b) match well with those in the experimental spectrum in figure \ref{figlabel_3}(a), with regard to the peak shape and energies;  the strong peak at +1.3 V in the experimental spectrum is the exception, and is exaggerated relative to PDOS by the asymmetry of the tunneling process. While PDOS of surface atoms will give some insight into STS, the extended states through the slab are also important, and we examined a series of energy windows, notated from I to V in figure \ref{figlabel_3}(b), by plotting charge density across the slab [figure \ref{figlabel_3}(c)].
The peaks found in windows I, II and III correspond to bonding states in the Si tetramer, which produce the protrusion in the topographic STM images. In particular, the three peaks in range III, situated in the band gap of bulk silicon, consist of three bands crossing the Fermi energy (further detail of the band structure can be seen in figure \ref{figlabel_A3}). Therefore, electronic transport under small bias voltages ($\textless\pm$0.25 V) will occur only within the FeSi$_2$ film. In energy window IV, the surface states are completely absent and the charge density is depleted from the first to fifth atomic layers below the surface. We also note the presence of significant charge density at the interface between the FeSi$_2$ film and the Si(001) substrate in ranges III and IV (see also figure \ref{figlabel_A3}); a detailed characterization of the atomic and electronic structure of the interface between Si(001) and the FeSi$_{2}$ film will be presented in future work. The onset of empty (anti-bonding) states of the Si tetramer occurs in energy window V.

\begin{figure*}
	\begin{center}
		\includegraphics[width=16cm]{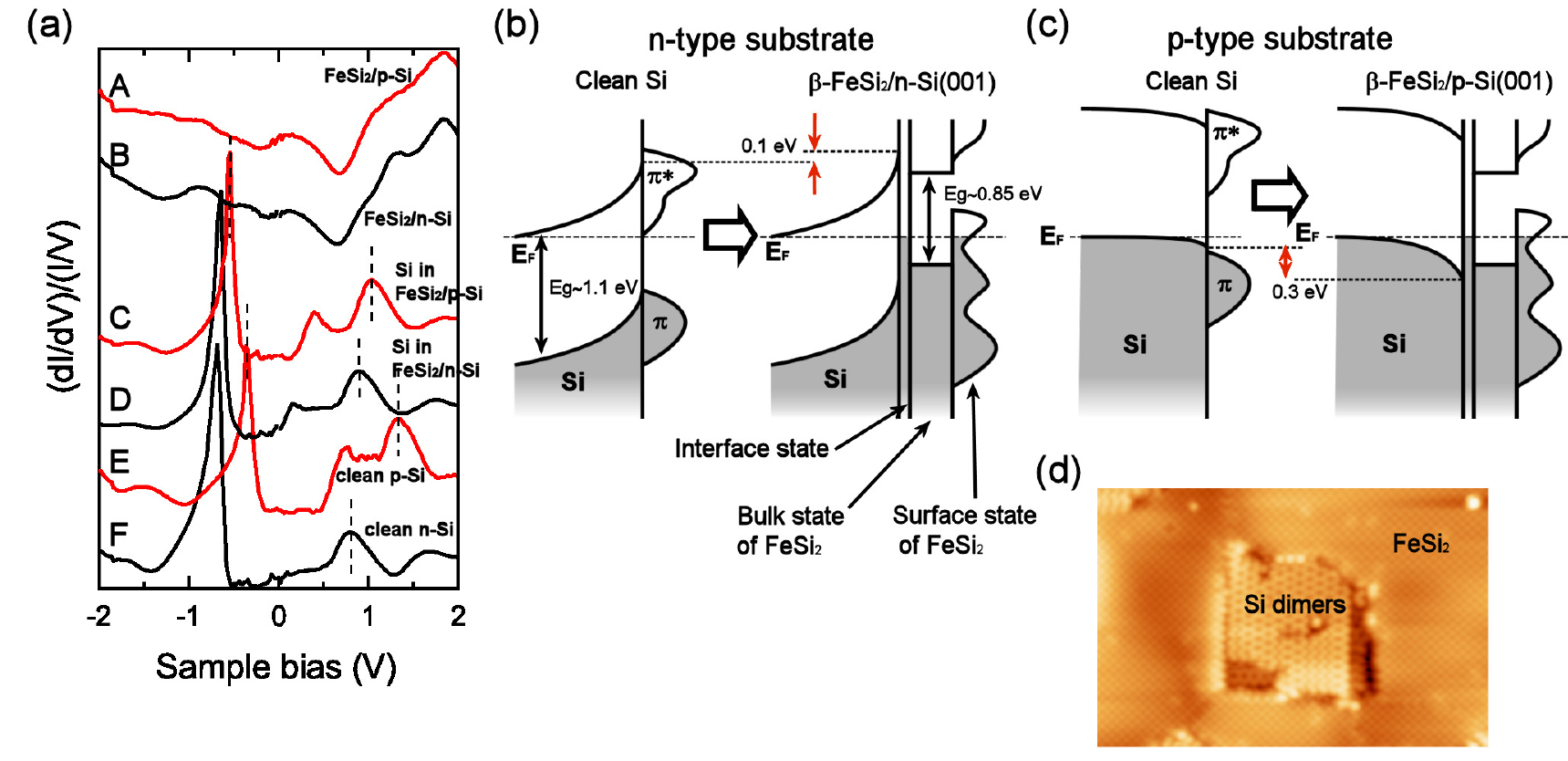}
	\end{center}
		\caption{(a) Normalized $dI/dV$ spectra recorded on $\beta$-FeSi$_2$ surface and Si dimers. Spectra A and B: $\beta$-FeSi$_2$ grown on p-type and n-type Si(001) substrates, Spectra C and D: Si dimers in a small domain not covered by $\beta$-FeSi$_2$ on p-type and n-type Si(001) substrates. Coverages of $\beta$-FeSi$_2$ on both samples were greater than 80\%. Spectra E and F: Si dimers on clean p-type and n-type Si(001) substrates without growth of $\beta$-FeSi$_2$. Set point: $V_s$= --2.0 V and $I$= 10 pA. Each spectrum was obtained by numerical differentiation after 16 I-V curves recorded in a 4 $\times$ 4 grid over an area of 1.5 $\times$ 1.5 nm$^2$ were averaged. (b) and (c) Band diagrams for clean Si(001) and $\beta$-FeSi$_2$/Si(001). (d) Representative STM image of a $\beta$-FeSi$_2$ film in the vicinity of a small Si dimer window.$V_s$ = +1.0 V, $I$ = 50 pA, $T_s$= 78 K.}
	\label{figlabel_4}
\end{figure*}

To our knowledge, the band alignment between $\beta$-FeSi$_2$ and the Si substrate has not been assessed in previous works. In order to clarify this point, we have examined the band bending in the cases of a $\beta$-FeSi$_2$ film grown on n-type and p-type Si(001) substrates. Figure \ref{figlabel_4}(a) shows normalized $dI/dV$ spectra recorded on the $\beta$-FeSi$_2$ surface and Si dimers on both n-type and p-type substrates at 78 K. Although the spectral features of $\beta$-FeSi$_2$ grown both on n- (Spectrum A) and p-type (Spectrum B) Si(001) were thermally broadened compared to those observed at 4.5K in figure \ref{figlabel_3}(a), the shapes and positions of the peaks in both spectra were almost equivalent. This indicates that the Fermi level of the FeSi$_2$ films is pinned at the same energy in the surface state on both the n- and p-type substrates, as depicted in figures \ref{figlabel_4}(b) and \ref{figlabel_4}(c). To achieve this, the electrostatic potential in Si near the interface with $\beta$-FeSi$_2$ must be bent in different manners for n- or p-type substrates. 

The evaluation of the band bending was performed by measuring the energetic positions of the Si dimer states in small windows of clean Si(001) in the $\beta$-FeSi$_2$ films [figure \ref{figlabel_4}(d)]. The Si dimer (Spectra C -- F) is characterized by a pronounced peak ($\pi$) in the occupied states and broad and intense peaks ($\pi_1$* and $\pi_2$*) in unoccupied states, separated by a surface band gap of 0.6$\sim$0.7 eV. \cite{Sagisaka05, Nakayama06}
The energetic locations of these states with respect to bulk valence and conduction bands have been recently reported. \cite{Sagisaka17} Based on that relationship, band diagrams for the clean Si surface are given for both n- and p-type substrates in figures \ref{figlabel_4}(b) and (c), respectively. \cite{Note1} By evaluating the peak positions, we found that the growth of FeSi$_2$ shifted the surface states of Si upward by 0.1 V on n-type Si [figure  \ref{figlabel_4}(b)] and downward by 0.3 V on p-type Si [figure \ref{figlabel_4}(c)]. Since we did not observe contrast abnormality in STM images, or a shift of spectral features in STS of the $\beta$-FeSi$_2$ surface in the vicinity of the uncovered areas, we believe that the same amount of band bending occurred both in the silicon underneath the $\beta$-FeSi$_2$ and in areas uncovered by the FeSi$_2$ film. This is reflected in the band diagrams, which imply that electrons (holes) are depleted in the n- (p-) type Si substrate near the interface after the formation of $\beta$-FeSi$_2$. 

The presence of a surface energy gap above the Fermi level affects the variation of the tunnelling conductance with bias voltage. To see this effect, we replotted the spectrum shown in figure \ref{figlabel_3}(a) in the form of I-V and dI/dV in figures \ref{figlabel_5}(a) and \ref{figlabel_5}(c), respectively. Above a sample bias ($V_S$) = +0.3, the current slightly drops, which is equivalent to the differential conductance becoming negative. Upon ramping up $V_S$, the current increases again (the differential conductance becomes positive). Similar NDC has been observed on various surfaces by STM \cite{Bedrossian89, Berthe08,Wang09,Kim13,Yin20}. For quantitative analysis, we computed an I-V curve from the DFT DOS based on the standard formula for the tunnel current:\cite{Hamers89}
\begin{equation}
	I \propto \int^{eV}_{0}\rho_s(E)\rho_t(eV-E)T(E,V,z)dE
\end{equation}
where $\rho_s$ and $\rho_t$ are the DOS of the sample and tip, and $T$ is the tunnelling matrix element. In the WKB approximation for planar electrodes, $T$ is written as $T(E, V, z) = \exp(-2z\sqrt{m(2\phi+eV-2E)}/\hbar)$, where $\phi$ is the effective work function. We used $\phi = (\phi_{tip} + \phi_{sample})/2$ = 9.4 eV and the tip-sample distance $z$ = 6 \AA\ for the computation of the I-V curve. Furthermore, we treated $\rho_{t}$ as a constant with respect to the energy, which simplified eq. (1) to:
\begin{equation}
	I \propto \int^{eV}_{0}\rho_s(E)T(E,V,z)dE
\end{equation}
For $\rho_s$, we used the Kohn-Sham eigenstates which also gave the PDOS shown in figure \ref{figlabel_3}(b). Computed I-V and dI/dV curves, shown in figures \ref{figlabel_5}(b) and \ref{figlabel_5}(d) respectively, successfully duplicate the overall peak positions and the feature of NDC of the experimental curves in figures \ref{figlabel_5}(a) and \ref{figlabel_5}(c). The NDC comes from the combined effect of the presence of an energy gap in the unoccupied state of the $\beta$-FeSi$_2$ surface and the behaviour of the tunnelling matrix $T$ with respect to energy, known as the tunnel-diode mechanism \cite{Bedrossian89} [figure \ref{figlabel_5}(e)]. 
The application of a small positive voltage to the sample induces tunnelling of electrons from the tip to the unoccupied state of the $\beta$-FeSi$_2$ surface [$V_S$ range I in figures \ref{figlabel_5}(b) and \ref{figlabel_5}(e)]. When the Fermi level of the tip ($E_{F, tip}$) matches the energy gap of the FeSi$_2$ surface, tunnelling of electrons from the tip to the surface is markedly suppressed, because $T$ is largest at $E_{F, tip}$ ($V_S$ range II). Consequently, the current decreases with increasing sample bias, while $E_{F, tip}$ is in the energy gap. As $E_{F, tip}$ exceeds the energy gap, the current increases again ($V_S$ range III). The excellent agreement between the experiment and simulation indicates that the NDC observed on the FeSi$_2$ surface is intrinsic to its electronic nature and is independent of that of the tip.

\begin{figure}
	\begin{center}
	\includegraphics[width=8.5cm]{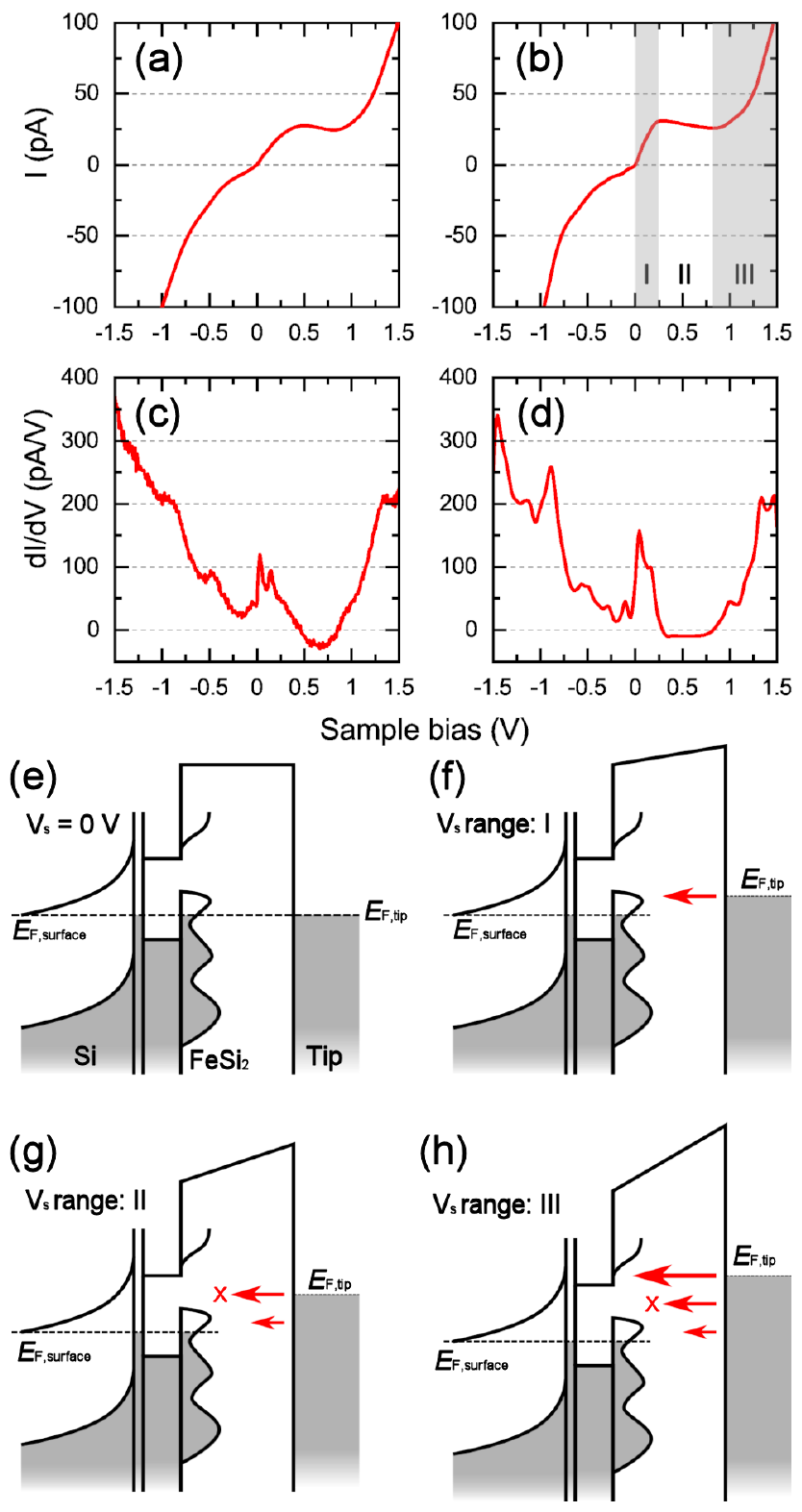}
	\end{center}
	\caption{Experimental and calculated tunnelling spectra from $\beta$-FeSi$_2$ surface. Experimental I-V curve (a) and dI/dV curve (c), and calculated DFT I-V curve (b) and dI/dV curve (d). Set point: $V_s$= +1.5 V and $I$= 100 pA. The experimental dI/dV curve was acquired by a lock-in technique with bias modulation of 10 mV at 830 Hz. The sample temperature was 4.5 K. (e)-(h) Schematic band diagrams for tunnelling processes with $V_s$ = 0 V and three different ranges. $V_s$ range I, II, and III correspond to those in (b). The symbol X in (g) and (h) indicates that tunnelling of electrons is not allowed at the corresponding $V_s$ due to the lack of electronic states within the surface band gap.}
	\label{figlabel_5}
\end{figure}

\section{Summary}
We have demonstrated the metallic nature of the surface of ultrathin $\beta$-FeSi$_2$ films grown on the Si(001) substrate. The excellent agreement of simulated STM, LDOS and I-V results with our experimental data confirms that the iron-silicide film  in this study was $\beta$-FeSi$_2$, rather than another possible crystal structure. We also determined band alignment between $\beta$-FeSi$_2$ and n- or p-type Si(001) substrate from tunnelling spectroscopy. Finally, I-V curves show NDC, owing to the presence of an energy gap in the unoccupied state of the $\beta$-FeSi$_2$ surface.
The new data for the $\beta$-FeSi2 film grown on the Si(001) substrate presented in this paper will provide an important launch point for the development of optoelectronic devices using this silicide. Moreover, the identification of NDC arising from the gap in the silicide film may guide the design of new materials systems with this important characteristic.  As we have shown, the origin of the metallic surface state is the Si tetramers, so the formation of Si clusters in or on $\beta$-FeSi2 may cause unwanted electronic states in the band gap. This implies that growth of a high quality film, with the surface state removed, will be necessary for device development. Finally, the band alignment between $\beta$-FeSi2 and silicon presented in figure~\ref{figlabel_4} will form a good starting point for designing and controlling the bands in this material by impurity doping.

\ack{Computational calculations were performed by using the Numerical Materials Simulator of
NIMS. This study was partly supported by JSPS KAKENHI Grant Number 22K04860.}

\appendix

\section*{Appendix}
\setcounter{section}{1}

\begin{figure}[b]
	\begin{center}
		\includegraphics[width=6cm]{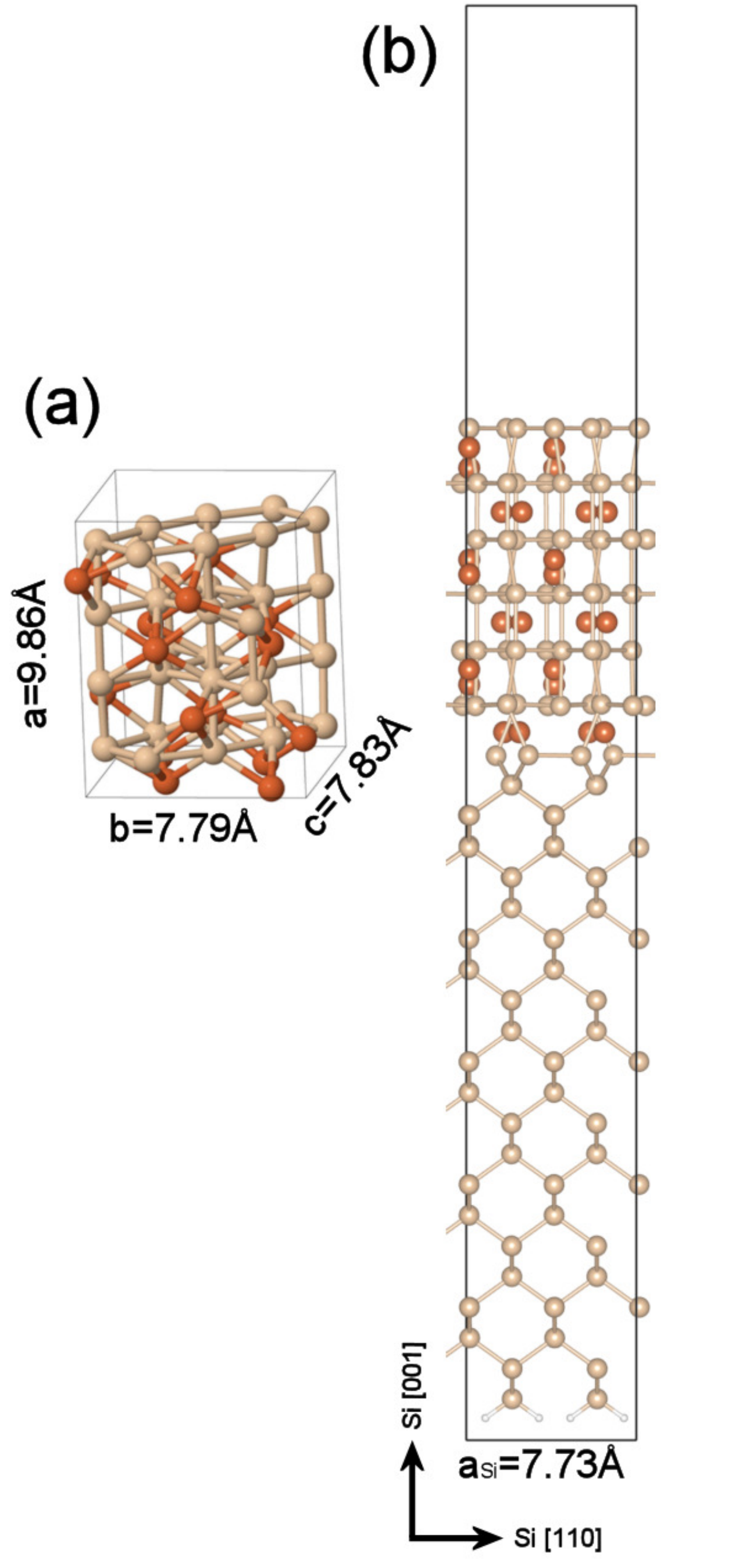}
	\end{center}
	\caption{(a) Unit cell of $\beta$-FeSi$_2$,\cite{Dusausoy71} (b) Slab model for the $\beta$-FeSi$_2$ film and Si(001) substrate.}
	\label{figlabel_A1}
\end{figure}

Figure~\ref{figlabel_A1}(a) shows the  shows the $\beta$-FeSi$_2$ unit cell, while Figure~\ref{figlabel_A1}(b) shows the slab model for the $\beta$-FeSi$_2$/Si(001) system that was used for calculations of STM images in figures \ref{figlabel_2}(b) and \ref{figlabel_2}(d), PDOS in figure \ref{figlabel_3}(b), charge density plots in figure \ref{figlabel_3}(c), and I-V curve in figure \ref{figlabel_5}(b). The structure for the interface between $\beta$-FeSi$_2$/Si(001) substrate was determined by an analysis of cross sectional TEM images\cite{Sagisaka20}.

\begin{figure}
	\begin{center}
		\includegraphics[width=13cm]{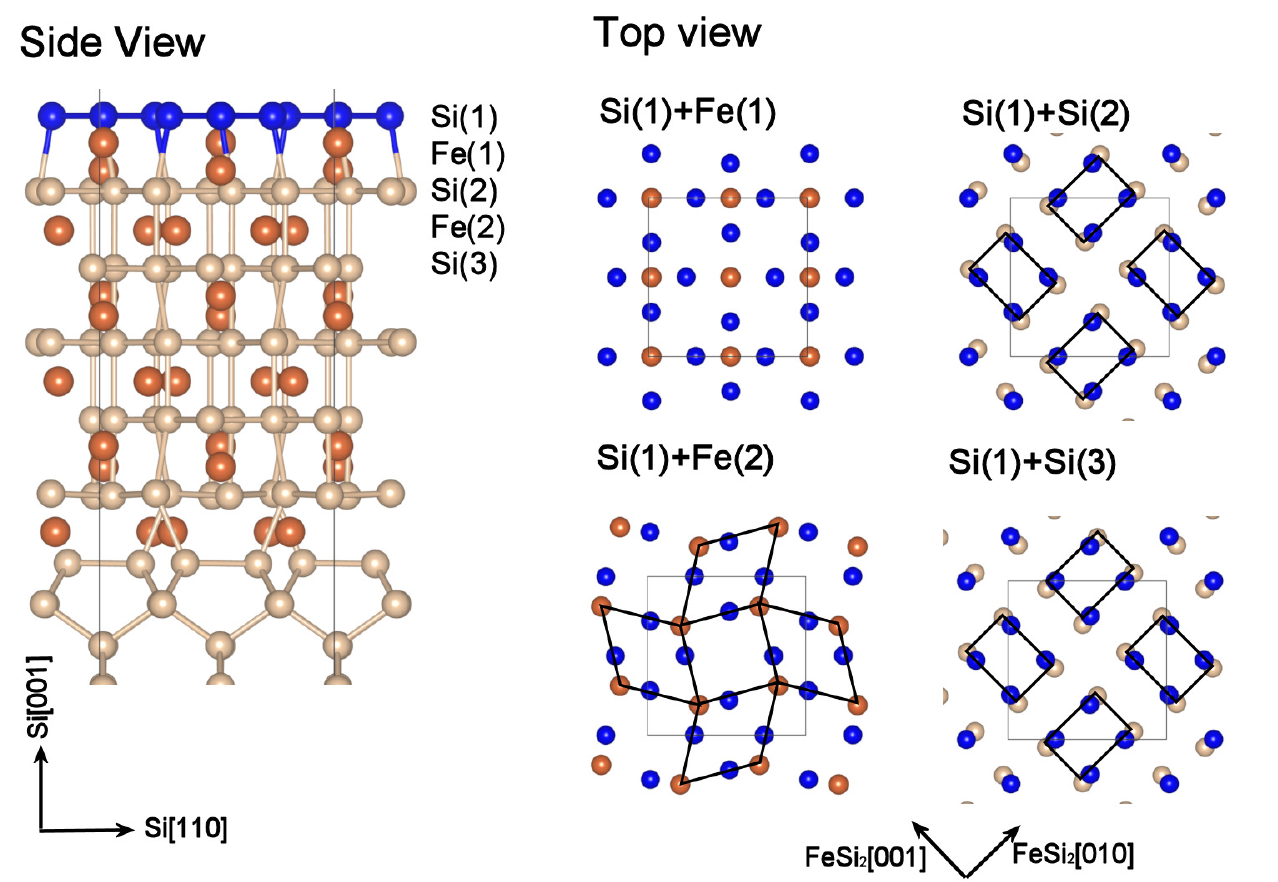}
	\end{center}
	\caption{Relative atom positions of subsurface layers in the $\beta$-FeSi$_2$ film with respect to the Si clusters in the first layer.}
	\label{figlabel_A2}
\end{figure}

In the main text, we mentioned the alternating elongated protrusions in the magnified STM image of the $\beta$-FeSi$_2$ surface [figure \ref{figlabel_2}]. The origin of this behaviour is probably derived from the alternation of the lattice symmetry of Si atoms and Fe atoms in subsurface layers, as displayed in figure \ref{figlabel_A2}, because the positions of the Si atoms in the topmost layer closely maintain four-fold symmetry. The Si atoms in the fourth layer [Si(2)] under the Si cluster in the top layer [Si(1)] are positioned in rectangles that alternately change their orientation between the FeSi$_2$ [001] and [010] crystal orientations. The Fe atoms in the fifth layer [Fe(2)] are positioned in diamonds that also alternately change their orientation.  

\begin{figure}
	\begin{center}
		\includegraphics[width=15cm]{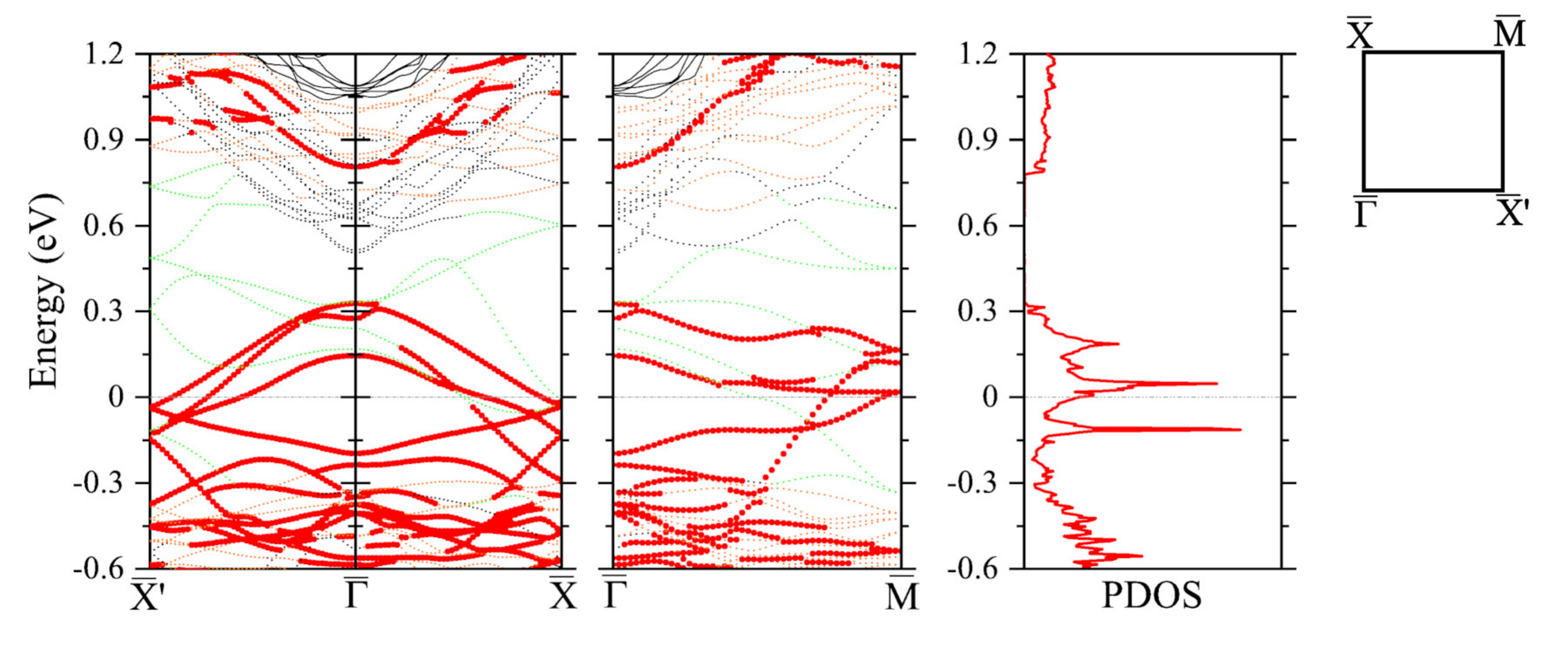}
	\end{center}
	\caption{Band diagram and PDOS calculated for the $\beta$-FeSi$_2$/Si(001) slab. Bands were projected in the surface BZ, (see~\cite{Romanyuk2014} for definitions of special points). The eigenvalues with significant charge ($\textgreater$0.001 e/\AA) on the Si atoms in the top layer are plotted with red dots, whereas those in the $\beta$-FeSi$_2$ film and Si slab are plotted with orange and black dots, respectively. Bands presented with green dots appear at the interface between the $\beta$-FeSi$_2$ film and Si(001) substrate.}
	\label{figlabel_A3}
\end{figure}

Figure \ref{figlabel_A3} shows electronic bands projected in the surface BZ and DOS projected on the topmost Si atom calculated with the slab model shown in figure \ref{figlabel_A1}. Although bands originating from the Si layers are included, overall features corresponding to the $\beta$-FeSi$_2$ film agree well with the previous report by Romanyuk \textit{et al.} \cite{Romanyuk2014} Compared to their results, the influence of in-plane compression in the silicide layer caused by lattice matching the Si(001) slab on the surface bands was not detected.

\begin{figure}
	\begin{center}
		\includegraphics[width=15cm]{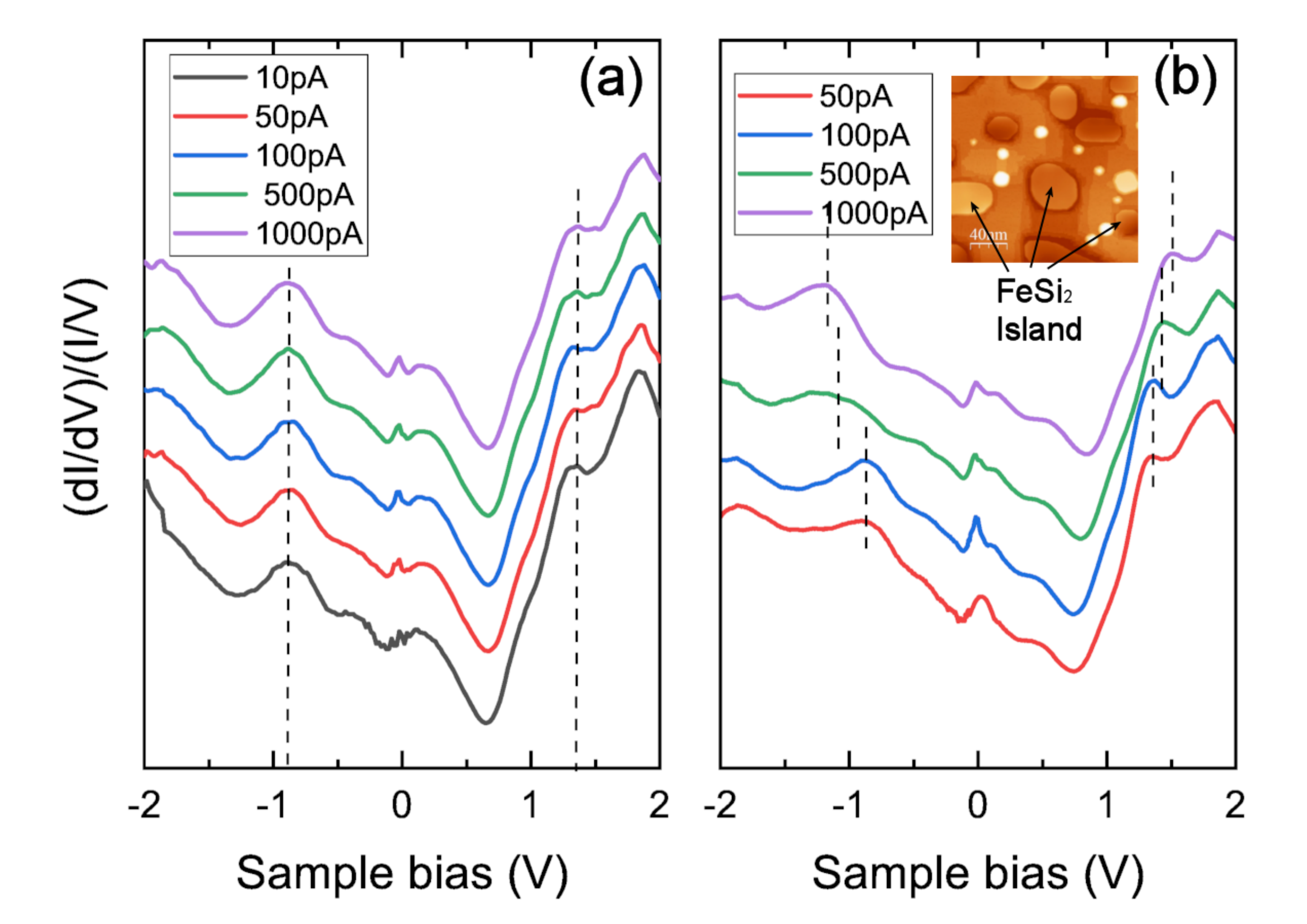}
	\end{center}
	\caption{Set point dependence of tunnelling spectra of the $\beta$-FeSi$_2$ surface. Spectra in (a) were acquired on the surface in figure \ref{figlabel_1} and spectra in (b) were acquired on one of $\beta$-FeSi$_2$ islands in the inset STM image. Set point: V$_s$ = +2.0 V, current is denoted in the graphs.}
	\label{figlabel_A4}
\end{figure}

The nature of the metallic electronic states of the $\beta$-FeSi$_2$ surface can be found in the result of STS measurements for different tip-surface distances. Figure \ref{figlabel_A4} shows normalized $dI/dV$ curves recorded with different set point currents. As the current increases by one order of magnitude, the tip-surface distance increasea by approximately 1 \AA. For a surface mostly covered by $\beta$-FeSi$_2$ (figure \ref{figlabel_1}), peaks in the spectra remain at the same sample bias, even when the tip-surface distance is changed [figure \ref{figlabel_A4}(a)]. This is ascribed to the screening effect of the surface on the strong electric field induced by the STM tip, indicative of the presence of a significant carrier density at the Fermi level. On the other hand, for a sample with a low coverage of $\beta$-FeSi$_2$ [inset in figure \ref{figlabel_A4}(b)], peak positions are shifted away from the Fermi level as the tip-surface distance is decreased [figure \ref{figlabel_A4}(b)]. Because of the low coverage of $\beta$-FeSi$_2$, screening does not work effectively and tip-induced band bending occurs. A similar behaviour can be seen on  the clean surface of Si(001) with the same doping level as used in this study [figure 3(a) in ref. 22].    

\pagebreak

\end{document}